\documentclass[letterpaper,twocolumn,10pt]{article}
\usepackage{usenix,url,amsfonts,amsmath,amsthm,multirow,amssymb,bmpsize}
\usepackage{graphicx}
\usepackage{enumitem}
\setlist{noitemsep}

\newcommand\mytt[1]{\texttt{\small{#1}}}

\newcommand\ak{$(\alpha,\kappa)$ }
\newcommand\aks[2]{$(\alpha={#1},\kappa={#2})$}

\begin{document}

\date{}

\title{\Large \bf Diffix-Birch: Extending Diffix-Aspen}

\author{
{\rm Paul Francis$^{\dag}$ \quad Sebastian Probst-Eide$^{\S}$ \quad Pawe\l{} Obrok$^{\S}$}\\
{\rm Cristian Berneanu$^{\S}$ \quad Sa\v{s}a Juri\'{c}$^{\S}$ \quad Reinhard Munz$^{\dag}$}\\
$^{\dag}$Max Planck Institute for Software Systems (MPI-SWS), Germany\\
$^{\S}$Aircloak GmbH, Germany\\
\{francis, munz\}@mpi-sws.org, \{sebastian, sasa, pawel, cristian\}@aircloak.com
}

\maketitle


\subsection*{Abstract}


A longstanding open problem is that of
how to get high quality statistics through direct
queries to databases containing information about individuals
without revealing information specific to those individuals.
Diffix is a framework for anonymous database query that
adds noise based on the filter conditions in the query.
A previous paper described the first version, called diffix-aspen.
This version, diffix-birch, extends that description to include
a wide variety of common features found in SQL. It describes
attacks associated with various features, and the anonymization
steps used to defend against those attacks.
This paper describes diffix-birch, which was used for the bounty
program sponsored by Aircloak starting December 2017.


\section{Introduction}
\label{sec:intro}

Diffix-birch\footnote{
    This is the second version of the ArXiv paper. It differs from
    the first version only in that it renames Diffix to diffix-birch.
} is the second version of
a new approach to anonymized database query that adds
noise to answers, but does so in a way that takes into account
the filter conditions of the query. In doing so, it minimizes the
amount of noise needed to strongly protect the anonymity
of individuals in the database, and eliminates the need for
the budget that is found in systems based on differential privacy.
A previous paper~\cite{diffix17} motivated and described the
first version, diffix-aspen, as applied to a
simple query language: one that allowed only the query conditions
\mytt{WHERE column = value} and \mytt{WHERE NOT column = value} in a
simple SQL \mytt{SELECT}.

Diffix-birch, described in this paper,
extends the query semantics in~\cite{diffix17} to include 
a wide variety of SQL features including sub-queries, \mytt{JOIN}, 
\mytt{GROUP BY} and \mytt{HAVING}, \mytt{LIKE}, \mytt{IN},
and a variety of math, string, and 
datetime functions. In so doing, diffix-birch substantially increases
the utility of the system, but also substantially increases the
size of the attack vector.
This paper describes many new attacks that are possible because of
the expanded semantics, and the subsequent defenses.

For the most part, this paper does not require knowledge
of~\cite{diffix17}.  On occasion this paper
uses text from~\cite{diffix17} without attribution.

Note that diffix-birch is the version that was used
for the bounty program sponsored by Aircloak and described at
challenge.aircloak.com. This paper was previously distributed 
at that URL in December of 2017. This ArXiv reference may
in particular be used by authors who wish to publish their attacks.

\section{A brief history of anonymization}
\label{sec:background}

As early as the mid-1800's, confidentiality of individuals in
the U.S. census became a concern~\cite{censusHistory}. The census bureau for
instance started
removing Personally Identifying Information (PII) like names
and addresses from publicly available census data. Over the
ensuing decades, the bureau increasingly used a variety of
techniques to mitigate the possibility that micro-data or tabulated data
would allow individuals in the data to be identified.
These techniques include rounding, adding random noise to values,
aggregation, cell suppression, cell swapping, and sampling
among others~\cite{censusHistory}.

In the 1950's, the bureau started using computers to tabulate data,
and by the 1960's anonymization techniques like those described above
were being automated~\cite{censusHistory}.
Computers introduced the ability for analysts to "cross-tabulate"
data (set filter conditions on queries). This tremendously increased
an analyst's ability to analyze the data, but also opened the
possibility that an analyst could \emph{isolate} an individual by specifying
a set of query conditions that uniquely identify that individual.

For instance, suppose that an analyst happens to know the birthday,
zip code, and gender of someone (the \emph{victim}).
Using SQL as our working query
language, the analyst could generate for instance the following query:

\vspace{0.3cm}
\begin{footnotesize}
  \begin{tabular}{l }
SELECT count(*) \\
FROM table \\
WHERE bday = '1994-02-05' AND \\
\hspace{1.0cm} gender = 'M' AND \\
\hspace{1.0cm} zip = 12345 \\
  \end{tabular}
\end{footnotesize}
\vspace{0.3cm}

If the answer is 1, then the analyst knows that the victim
is indeed the only individual in the dataset with those attributes.
Given this, the analyst can learn anything else about the individual.
For instance, to learn salary the analyst could make the following
query:

\vspace{0.3cm}
\begin{footnotesize}
  \begin{tabular}{l }
SELECT salary, count(*) \\
FROM table \\
WHERE bday = '1994-02-05' AND \\
\hspace{1.0cm} gender = 'M' AND \\
\hspace{1.0cm} zip = 12345 \\
\hspace{1.0cm} GROUP BY salary
  \end{tabular}
\end{footnotesize}
\vspace{0.3cm}

which would return the salary of the victim.

The basic defense against this, deployed in the 1960's, was to set
thresholds which required that a certain number $K$ of individuals
must be present in aggregated data for the aggregate to be released.
This $K$-threshold mechanism, however, does not prevent individuals
from being isolated.

\begin{figure}
\center
\begin{footnotesize}
  \begin{tabular}{| l | l |}
\hline
Intersection Attack & Section~\ref{sec:background} \\
Averaging Attack & Sections~\ref{sec:background}, \ref{sec:in} \\
Chaff Attack & Section~\ref{sec:background}, \ref{sec:chaff}, \ref{sec:posRange}, \ref{sec:in}, \ref{sec:like}, \ref{sec:upperLower} \\
Equations Attack & Sections~\ref{sec:background},~\ref{sec:eqAttack} \\
Split Averaging Attack & Section~\ref{sec:splitAv} \\
Difference Attack & Section~\ref{sec:stick}, \ref{sec:in}, \ref{sec:notin}, \ref{sec:like} \\
Backdoor Attack & Section~\ref{sec:details} \\
\hline
  \end{tabular}
\end{footnotesize}
\caption{Summary of Attacks
}
\label{fig:summary}
\end{figure}

The central problem is the \emph{intersection attack}.  By way of example, 
imagine a database that only returns answers that pertain to more
than $K=4$ individuals.  Suppose
that an analyst makes two queries, one for the number of people in 
the CS department, and one for the number of men in the CS
department:

\vspace{0.3cm}
\begin{footnotesize}
  \begin{tabular}{l | l}
SELECT count(*) &
SELECT count(*) \\
FROM table &
FROM table \\
WHERE dept = 'CS' AND &
WHERE dept = 'CS' \\
\hspace{1.0cm} gender = 'M' & \\
  \end{tabular}
\end{footnotesize}
\vspace{0.3cm}

Suppose that there are 34
people in the CS department and 33 of them are men.
Since both of these numbers
are greater than $K=4$, the database returns the answers.  The analyst
can trivially conclude that there is one woman in the CS
department even though the database would have refused to provide
that answer directly.  Armed with this knowledge, the analyst can then
learn more about the woman.  For instance, the analyst
can query for the sum of salaries of all people in the CS department
and the sum of salaries of all men in the CS department, and by
taking the difference determine the salary of the woman:

\vspace{0.3cm}
\begin{footnotesize}
  \begin{tabular}{l | l}
SELECT sum(salary) &
SELECT sum(salary) \\
FROM table &
FROM table \\
WHERE dept = 'CS' AND &
WHERE dept = 'CS' \\
\hspace{1.0cm} gender = 'M' & \\
  \end{tabular}
\end{footnotesize}
\vspace{0.3cm}

The earliest publication we could find that identifies the
intersection attack is by the statistician
Fellegi in 1972~\cite{fellegi1972question}.
In 1979, it was shown that even an analyst with no 
prior knowledge about the contents
of a $K$-threshold database that gives exact answers
can infer substantial information about individual users using
the intersection attack~\cite{denning1979tracker}.
It was also shown that one way to prevent the intersection
attack is to distort answers unpredictably,
for instance by randomly rounding user counts up or down to a
value divisible by five~\cite{fellegi1972question,fellegi1974statistical}
or removing random rows from the set of rows returned by the database
~\cite{denning1980secure}.

One problem with this approach arises if the analyst has the ability
to make an unlimited number of queries to the database.
If so, the analyst can remove the noise through
averaging: causing a given answer to be repeated, each
with a new noise sample, and taking the
average value.  We refer to this attack as the \emph{averaging attack}.

To defend against this, Denning et.al. proposed seeding the random
number generator with values taken from the query itself. The idea
here is that the same query would then produce the same noise.
We refer to this general concept of causing noise to
repeat as \emph{sticky noise}.
The problem with this particular sticky noise
solution, which indeed the authors recognized, is that the analyst can still
average out the noise by generating
multiple different queries that all produce the same result, for
instance by adding conditions that don't effect the answer
like \mytt{WHERE age < 1000}.
Systems that rely on interpretation of the query syntax, of which
diffix-birch is one, must defend against this sort of attack. We refer to
syntax changes that do not result in a change of answer as
\emph{chaff}, and refer to this kind of attack as a \emph{chaff attack}. 

Even in cases where answers cannot be repeated,
however, noise may be removed from data. In 2003, Dinur and Nissim
showed that the true values of cells in a database may be determined
with high confidence even where no query is
repeated~\cite{dinur2003revealing}. They do so by formulating a set
of queries where each query selects a different but overlapping
set of rows from the database, and computes a noisy sum.
They then generate a set of simultaneous equations from the queries,
and solve for the values.
We refer to this attack as an \emph{equations attack}.

Two big ideas in data anonymity research are K-anonymity~\cite{sweeney2002k}
and differential privacy~\cite{Dwork08}. Neither provides enough
utility to be generally and practically usable.
The idea behind
K-anonymity is that data values are aggregated to the point where
any given combination of attributes has at least $K$ distinct
individuals sharing the attribute values. Thus any given individual
is ``hidden'' in a group of at least K individuals. The problem
is that applying K-anonymity to all columns destroys the value
of the data~\cite{curse}. Applying K-anonymity to only
some columns (i.e. those that can be used to identify the individual),
as envisioned by Sweeney, still leaves much data unprotected.

Differential privacy, in a nutshell, protects data by adding
random noise. Ultimately
it defends against averaging and equations attacks
by limiting the number of noisy answers that may be reported to
an analyst.  This noise \emph{budget} severely limits the utility
of differential privacy. The only two operational deployments
of differential privacy that we are aware of, by
Google~\cite{erlingsson2014rappor} and Apple~\cite{appleDP},
both effectively operate with unbounded privacy loss.

\section{Definition and measure of anonymity}
\label{sec:anon}

The EU Article 29 opinion on anonymity~\cite{article29},
which serves as the basis for
evaluating anonymity in the EU GDPR (General Data Protection Regulation),
defines three essential risks to anonymization: \emph{singling out},
\emph{linkability}, and \emph{inference}. Article 29 defines these
three risks as follows:

\begin{quote}
{\bf Singling out}: which corresponds to the possibility to isolate some or all records which identify an individual in the dataset;

{\bf Linkability}: which is the ability to link, at least, two records concerning the same data subject or a group of data subjects (either in the same database or in two different databases). If an attacker can establish (e.g. by means of correlation analysis) that two records are assigned to a same group of individuals but cannot single out individuals in this group, the technique provides resistance against ``singling out'' but not against linkability;

{\bf Inference}: which is the possibility to deduce, with significant probability, the value of an attribute from the values of a set of other attributes.
\end{quote}

\begin{figure}[tp]
\begin{center}
\includegraphics[width=0.45\textwidth]{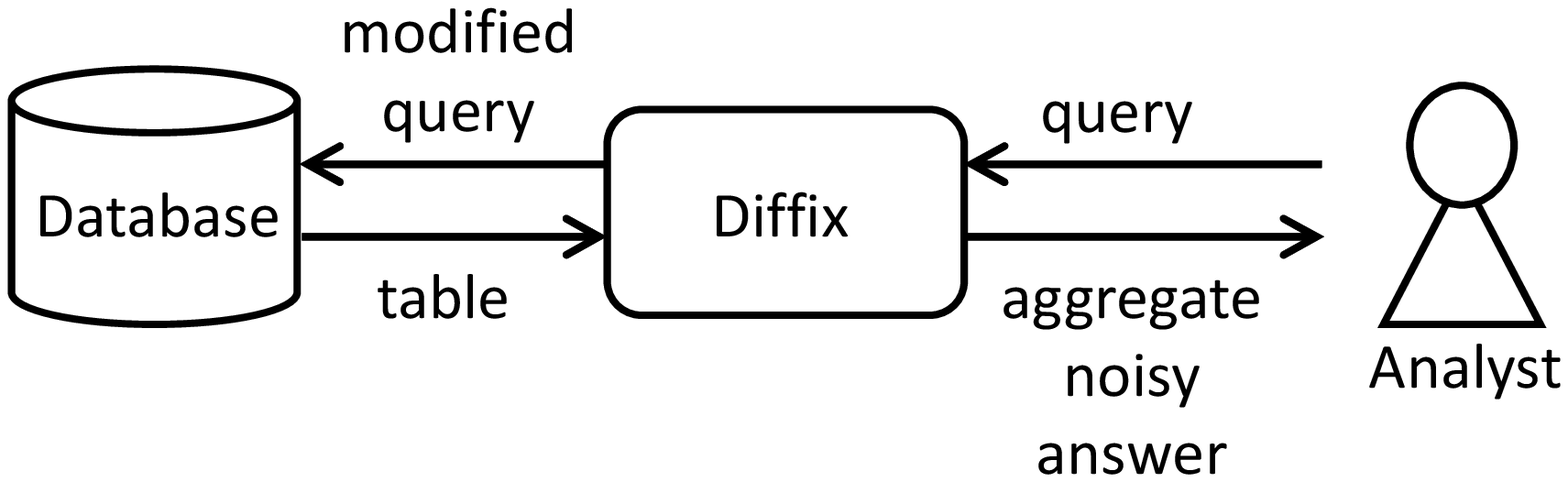}
\caption{Diffix-birch acts as an SQL proxy to an unmodified database}
\label{fig:system}
\end{center}
\end{figure}

We base our definition of anonymity on these risks, not just because these
definitions are used by practitioners, but also because 1) we find them to make
sense intuitively, and 2) we can formulate tests that measure the risks.
These definitions, however,
imply an anonymization system that retains the notion of a
``record''.  Diffix-birch does not anonymize the dataset per se, but rather
anonymizes answers to queries on the dataset: the analyst does not
have direct access to the dataset and its records (Figure~\ref{fig:system}).
We therefore modify these definitions to fit diffix-birch's operational
model as follows.

We define \emph{singling out} as occurring when an analyst 
correctly makes a statement of
the form ``There is exactly one user that has these attributes.''
For instance, the analyst may \emph{claim} that there is a single user
with attributes [gender = 'male', age = 48, zipcode = 48828,
lastname = 'Ng']. If this is true, then the analyst has correctly
singled out that user.  The attributes don’t need to be personal
attributes as in this example.  If the analyst correctly claims
that there is a single person with the geo-location attributes [long = 44.4401,
lat = 7.7491, time = '2016-11-28 17:14:22'], then that person is
singled out.

We define \emph{linkability}
in the context of a second dataset
(the \emph{linkability dataset})
that has some users in common with the dataset
behind diffix-birch (the \emph{protected dataset}).
In other words, for some fraction of rows in
the linkability dataset, there are rows in the protected dataset that
belong to the same user.

Assuming that the analyst has full access to the linkability dataset,
linkability occurs when an analyst correctly makes a statement
of the form ``this user or set of users in the linkability
dataset also exist in the protected dataset.''

We define \emph{inference} in the context of a second
dataset (the \emph{inference dataset}) which is identical to
the protected dataset, but where one or more cells in the
inference dataset are masked (the values are unknown to
the analyst). Other than this masking, the analyst has
full access to the inference dataset.

Inference occurs when an analyst correctly makes a statement
of the form ``the value of this masked cell in the
inference dataset is X''.

Of course, some fraction of an analyst's claims
may be incorrect. This leads to the
notion of \emph{confidence}, which is defined as the ratio of
correct claims to all claims.  If 95\% of an analyst’s claims are
correct, then the confidence of the attack is 95\%. \emph{Confidence
improvement} (or just \emph{improvement} for short) is the improvement in
the analyst’s confidence over a statistical guess.

By way of example,
suppose that roughly 90\% of the users in the database have zipcode = 48828,
and that the analyst knows this.
Suppose also that the analyst has external knowledge that there is
a single user with the
attributes [gender = 'male', age = 48, lastname = 'Ng'].
The analyst could simply claim that this user’s zipcode = 48828,
and the analyst would have 90\% confidence in the claim.
But this does not improve on a statistical guess, and so
Diffix-birch has not leaked additional information with respect to the
inference. The improvement is therefore zero.

We measure confidence
improvement as $\kappa = (C-S)/(1-S)$, where $C$ is the analyst’s
confidence, and $S$ is the statistical probability.
So for example if the analyst were, through some attack on diffix-birch,
to improve their confidence from 90\% to 95\%, then the
confidence improvement would be $\kappa=0.5$ (or 50\%).

The inference and linkability datasets imply some amount of
externally derived \emph{prior knowledge}.
Our measure of the quality of diffix-birch's anonymity accounts for
this prior knowledge. 
Prior knowledge may be gained from external data
sources, or the analyst may literally know portions of the
database, for instance because a database that the analyst has
full access to has been joined with other databases to form
the protected database.

We define $P$ as the number of cells an analyst knows as prior
knowledge, where a cell is a single value in the database (the
value at a single column and row). Of course a cell may itself
be a complex object with multiple values, but for our purposes
we don't worry about that. For singling out and inference attacks,
we define $L$ as the number of cells an analyst learns. 
For linkability attacks, we define $L$ as the number of row
linkages the analyst learns.

This allows us to define $\alpha=L/(P+1)$
as a value that expresses the strength of diffix-birch's
anonymization (or that of any anonymization scheme) with respect
to the analyst's prior knowledge.
1 is added to the denominator to avoid divide-by-zero.

Of course when an analyst ``learns'' something, it is with a
certain confidence improvement $\kappa$. Therefore, we characterize the
strength of an anonymization scheme with both
parameters $(\alpha,\kappa)$. This strikes us as a reasonable though
still flawed
way for a privacy stakeholder, for instance a Data Protection Officer (DPO),
to think about the value of an anonymization scheme. For instance,
suppose that a given system has \aks{10e-6}{0.9}.
This means that an attacker must know a million things in order to
learn one thing with a 90\% confidence improvement. The DPO
might or might not regard this as an acceptable risk, but at least
the DPO will have some reasonable sense of the risk.

The reality of course is more complex. It may well be that knowing
certain cells makes the system more attackable than knowing other
cells. For instance, knowing all of a column that on its own
isolates many users may be better for an attacker than knowing
all of a column that isolates no users. It may also be that, given
certain prior knowledge, certain cells are easier to learn than others.
The sensitivity of the easier-learned cells may also be a factor.

What's more, an \ak score can only really be measured with respect
to some specific attack. A DPO might therefore know be told
something like ``if an analyst knows columns A and B, and does a
foobar attack, then he can learn things about column C with
\aks{10e-7}{0.85}''.

Note that our immediate purpose for defining \ak is in order to
assign monetary payoffs in a bug bounty challenge. With respect to
this paper at least, \ak conveys a sense of how we think about
measuring anonymity.

\subsection{Limitations on GDPR anonymity criteria}

{\bf Singling out:}
By our definition, a user is singled out even if an analyst makes
a claim about a single cell value, as long as that value isolates the user
and is not prior knowledge. There are cases, however, where 
the analyst may exploit specific patterns in the column data to
make successful claims without in fact violating privacy in any
meaningful way.

Common among these is the case where uid's are assigned sequentially.
Since the analyst knows that each uid is distinct, it is quite
easy using diffix-birch to determine that the uid's have been sequentially
assigned, and roughly what the low and high uid values are.
Given this, the analyst could make a series of singling out claims
that there is one user with uid = 1, one user with uid = 2, and so on.
Intuitively, we regard this as not violating privacy because it
tells us nothing specific about the singled out user---one user
could have a given uid as easily as the next.

We have no precise definition of when exploiting
such predictable patterns in data are and are not privacy violating.
For now we have to take these on a case-by-case basis.

{\bf Exposure of strings (security versus privacy):}

It may be a security violation that any string in a database is revealed.
For instance, if a column contains passwords, then it would likely be
a security violation that any password is revealed, even if many
users happen to have the same password. Neither diffix-birch nor
the GDPR criteria for anonymity account for this requirement. If
such a column exists in a database, then it must be removed or made
inaccessible by diffix-birch.

\section{Assumptions and terminology}
\label{sec:assume}

Our system setup consists of an \emph{analyst} that queries a \emph{database}
via diffix-birch (Figure~\ref{fig:system}).
The database is conceptually a single table organized as rows and columns.
The columns may be of any type, so long as there are equalities
and inequalities defined for the type (e.g. \emph{column = value} or
\emph{column $<$ value}) returning TRUE or FALSE.
The database holds ``raw'' data: no perturbation on the values in
the database is assumed, and no columns need be removed\footnote{
The exception is columns that are a security risk as opposed to a
privacy risk, as with the password column example.
}, for instance
those containing personally identifying information like names.

We refer to the entity whose privacy is being protected as the
\emph{user}.  The user may well be a device like a smartphone or
a vehicle or even an organization.
We require that each database table with individual user data
has a column containing user identifiers.
This is typically nothing more than the Primary Key or Foreign
Key in the relational database.
By convention we call this column the \emph{uid}.  We assume that
every distinct user has one and only one distinct uid.
A user may of course have more than one row.

The database may change over time.  However, to protect anonymity
in the face of changes, all changes to the database must be
timestamped, and all queries must have a time range associated
with the query.  Note that as of this writing, our implementation of
Diffix-birch does not have mechanisms to ensure this.

Diffix-birch must be configured to know 1) which column contains the timestamps,
and 2) which column contains the uid's.
No other data-specific configuration is required.
Critically, the system operator need not understand the semantics
or sensitivity of any other columns.

This paper assumes that each database table has only a single uid column,
and that any given row contains information only about the user
identified by the uid.  This precludes certain data structures for
social network data.  For instance a row identifying both 
the sender and recipient of a message is disallowed.  Rather,
such a message would have to be encoded as two rows, one for the
sender and one for the recipient.

Diffix-birch has no specific mechanisms to deal with perfect
correlations between columns among groups of users.
If such correlations exist, and there is a risk that an analyst
knows of the correlation, then the amount of noise
must be increased in proportion to the size of the correlated group.
For instance, if 10 people in a geolocation database
travel together as a group, then to protect the privacy of all users in
the group the noise must be increased 10x.
The noise amount is statically configured and applies to all answers.

An analyst may make an unlimited number of queries.  

This paper does not address timing or other side-channel attacks.

\section{Basic Design}
\label{sec:basedesign}

A \emph{query} may have
zero or more \emph{filter conditions} (or just conditions).
The conditions determine which rows of the database comprise
a given answer.  A query may have one or more
\emph{anonymizing aggregation functions}. In diffix-birch,
these include \mytt{count}, \mytt{sum},
\mytt{avg}, \mytt{stddev}, \mytt{min}, \mytt{max}, and \mytt{median}.

The query produces a
response that consists of one or more \emph{columns}, and
zero or more \emph{rows} or \emph{buckets}.
For example, the following query has two conditions (\mytt{dept}
and \mytt{salary}), and one anonymizing aggregation function
(\mytt{count(*)}). The response has two columns (\mytt{salary}
and \mytt{count(*)}), and multiple buckets (response rows), where each
bucket corresponds to a given salary.
The \mytt{dept} condition excludes any database rows
that don't have a department of 'CS', and the \mytt{salary}
condition excludes any database rows from each bucket that do not have
the salary corresponding to that bucket.

\vspace{0.3cm}
\begin{footnotesize}
  \begin{tabular}{l}
SELECT salary, count(*) \\
FROM hrtable \\
WHERE dept = 'CS' \\
GROUP BY salary \\
  \end{tabular}
\end{footnotesize}
\vspace{0.3cm}

Diffix-birch distorts responses it two ways:
\begin{enumerate}
\item It may change the aggregation function's true value.
\item It may suppress buckets.
\end{enumerate}

\begin{figure}[tp]
\begin{center}
\includegraphics[width=0.42\textwidth]{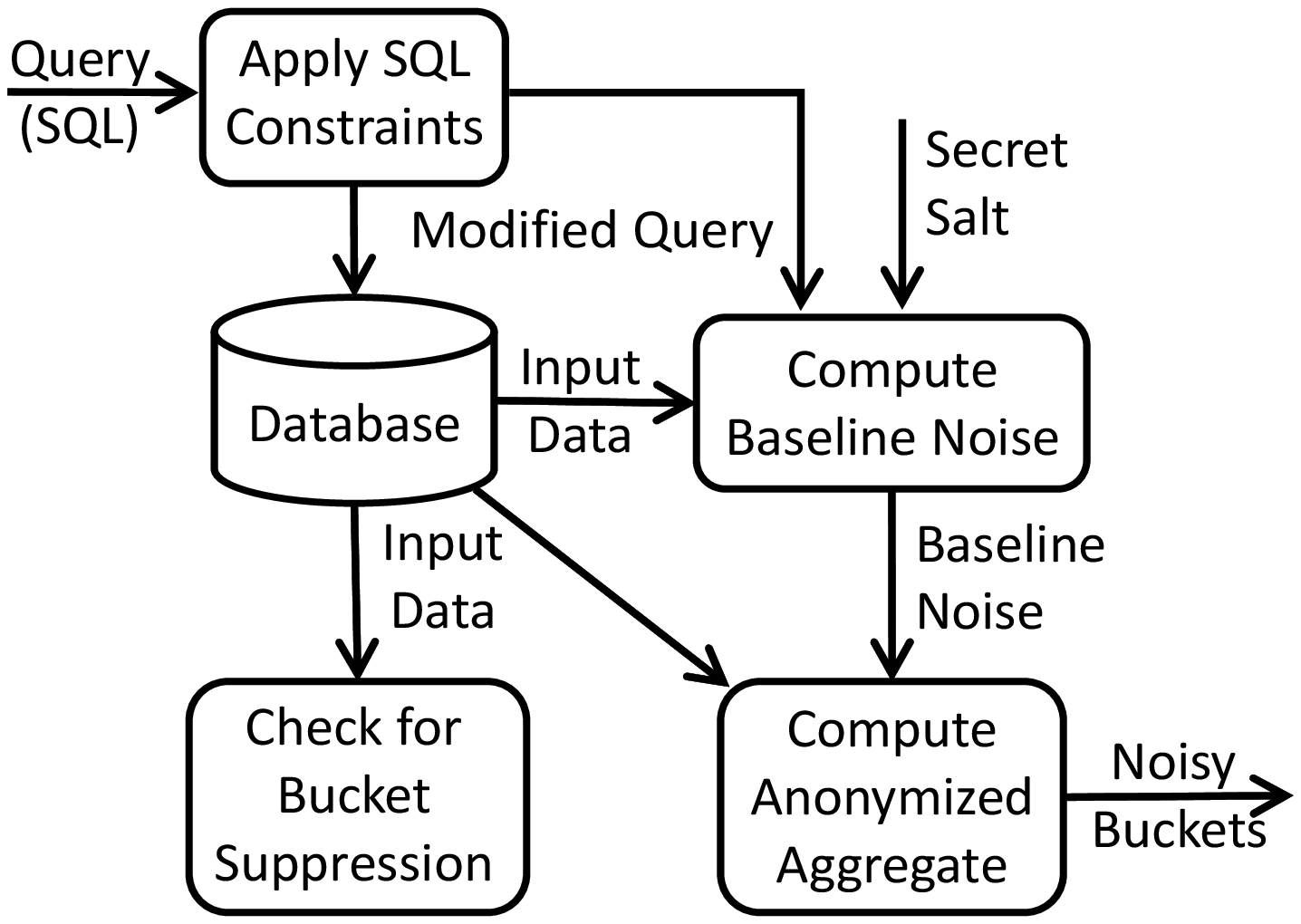}
\caption{An analyst query selects data from a database based on
filter conditions in the query. The database output serves as
the \emph{input data} to an anonymizing aggregation function.
Noise is computed based on input from both the query and
the input data.}
\label{fig:baseflow}
\end{center}
\end{figure}

The change in aggregation function value is normally due in part to added
pseudo-random noise taken from a Gaussian distribution. It may
also in part be because of other mechanisms such as removing outliers
from the input data (where input data here refers to the input of
the aggregation function, see Figure~\ref{fig:baseflow}).
Bucket suppression happens when the input data for any given bucket
has too few distinct users.
For example, a bucket for the salary \$100K that has a true value of 20
users may report a noisy value of 22 users.
A bucket for the salary \$450K that has only a single user will
be completely suppressed.

\subsection{Sticky Layered Noise}
\label{sec:stick}

A key concept in diffix-birch is that of \emph{sticky layered noise}.
Diffix-birch's noise is layered in that a bucket is not distorted
by a single noise sample, but rather by the sum of
multiple noise samples, where each sample is related to a condition.

Diffix-birch controls how a noise sample is generated by how it seeds the
Pseudo-Random Noise Generator (PRNG): the same seed produces the
same noise sample.
Exactly how a seed is produced depends on the type of condition, but the
general idea is to take aspects of the condition itself (the column
name, the value, the operator), combine them with
a secret salt, and use that as the seed.

In fact diffix-birch has two ways of seeding, or said differently, diffix-birch
has \emph{two types of noise}. We refer to these as \emph{static}
and \emph{dynamic} noise. Static noise defends against \emph{averaging
attacks}, and dynamic noise defends against \emph{difference attacks}
(described further in this section).

As an example, the seed for the static noise layer associated
with the condition \mytt{WHERE dept='CS'} in the above query
is generated as\footnote{This is slightly simplified, see
Section~\ref{sec:posEq}.}:

\begin{footnotesize}
\begin{verbatim}
static_seed =
   XOR(hash(concat('hrtable','dept','CS')),salt)
\end{verbatim}
\end{footnotesize}

The difference between static and dynamic noise layers is that
dynamic noise layers \emph{additionally} include the distinct set
of \mytt{uid}'s for the users whose rows comprise a given bucket.
As such, the seed for the
dynamic noise layer for the same condition would be:

\begin{footnotesize}
\begin{verbatim}
dynamic_seed =
   XOR(static_seed,hash(uid1),hash(uid2),...)
\end{verbatim}
\end{footnotesize}

Thus static noise layers are static in the sense that the same
condition generates the same noise no matter what query it
appears in, whereas a dynamic noise layer for a given condition
will differ as long as the set of users in the bucket differ.
For example, the static noise layers for \mytt{dept='CS'} in the
following two queries are the same, while the dynamic noise layers will
differ from each other so long as there are CS members younger
than 30 or older than 40.

\vspace{0.3cm}
\begin{footnotesize}
  \begin{tabular}{l | l}
SELECT count(*) &
SELECT count(*) \\
FROM hrtable &
FROM hrtable \\
WHERE dept = 'CS' AND &
WHERE dept = 'CS' \\
\hspace{1.0cm} age BETWEEN 30 AND 40 & \\
  \end{tabular}
\end{footnotesize}
\vspace{0.3cm}

For queries that count distinct users, the standard deviation of
each noise layer is $\sigma=1$. As a result, a single noise layer
taken alone introduces a fair amount of uncertainty as to the exact
value of a distinct user count, and multiple noise layers even more so.

\subsection{Attacks}

With sticky layered noise, the
simple averaging attack of Section~\ref{sec:background} doesn't
work because the noise doesn't change with each repeated
bucket. The following sections explore additional attacks.

\subsubsection{Split Averaging Attack}
\label{sec:splitAv}

At this point one might suppose that noise needs to be sticky
to prevent the averaging attack, but not necessarily layered.
Here we describe a more sophisticated variant of the averaging
attack, called the \emph{split averaging attack} that justifies
the need for layers.

In the split averaging attack,
the analyst produces the pair of queries shown in Figure~\ref{fig:split}.

\begin{figure}
\center
\begin{footnotesize}
  \begin{tabular}{l | l}
\hline
SELECT count(*) &
SELECT count(*) \\

FROM table &
FROM table \\

WHERE dept = 'CS' AND &
WHERE dept = 'CS' AND \\

\hspace{1.0cm} age = 20 &
\hspace{1.0cm} age $<>$ 20 \\
\hline
  \end{tabular}
\end{footnotesize}
\caption{One pair of queries from a split averaging attack to learn exact count of people in the CS dept. Subsequent query pairs use \mytt{age=21,22,...}}
\label{fig:split}
\end{figure}

The sum of the counts of the two queries gives the number of
users in the CS department, plus noise (here assuming that there
is one row per distinct \mytt{uid}).
Now repeat the pair of queries, this time using \mytt{age=21} and
\mytt{age<>21}. This produces the same sum, but with different
queries.  The pairs can be repeated with \mytt{age=22,23,24,}
and so on. 

If noise were sticky but not layered, then each individual query would have
a different noise sample because each query is different.
With enough samples, the noise could be averaged away and
a high-confidence exact count produced. Given exact counts, the
analyst could then for instance carry out an intersection
attack~\cite{denning1979tracker}.

With static noise layers, however, the attack doesn't work.
Each query has two static noise values:
one based on the condition \mytt{dept='CS'},
and one based on the age condition (\mytt{age=XX} or \mytt{age<>XX}).
The \mytt{age}-based noise value changes with
each query, and so could be averaged away. The \mytt{dept='CS'}
static noise, however, is always the same and is not averaged away.
The final averaged count would be perturbed by the \mytt{dept='CS'}
static noise layer.  

\subsubsection{Difference Attack}

The split averaging attack explains the need for static noise layers,
but not for dynamic noise layers. Indeed in the split averaging
attack, all of the dynamic noise layers differ, even the one for
\mytt{dept='CS'}, because each bucket has a different uid set.
Therefore they can also be averaged away and so don't help defend
against the split averaging attack.
The following attack, called the \emph{difference attack},
justifies the need for dynamic noise layers.

For this attack, suppose that
an analyst happens to know that there is only a single woman in the
CS department.  Let's call her the victim.
The analyst could form the two queries shown in Figure~\ref{fig:diff}.

\begin{figure}
\center
\begin{footnotesize}
  \begin{tabular}{l | l}
\hline
SELECT salary, count(*) &
SELECT salary, count(*) \\

FROM table &
FROM table \\

WHERE dept = 'CS' AND &
WHERE dept = 'CS' \\

\hspace{1.0cm} gender = 'M'  &
GROUP BY salary \\

GROUP BY salary & \\
\hline
  \end{tabular}
\end{footnotesize}
\caption{Difference attack to learn the isolated woman's salary}
\label{fig:diff}
\end{figure}

Both queries produce a histogram of salary counts.
The left query definitely excludes the victim from all answers.  The
right includes the victim only in the bucket that matches her salary.
We refer to this as \emph{isolating} the victim.
Suppose we take the difference in the count between each bucket
pair (two buckets representing the same salary).
If we only had static noise layers, then this difference would be
the same for every pair of buckets
for a salary other than that of the victim.
This is because the only difference between each such pair would the
static noise layer for \mytt{gender='M'}, which is always the same.
However, the bucket pair representing the
victim's salary would have an additional difference: the count
of the victim herself. As a result, to discover the victim's
salary, the attacker only needs to observe which bucket pair
has a unique difference.

The dynamic noise layer defends against this attack. Because the
set of uid's differs between pairs of buckets, the difference
in noise between
each pair also differs. As a result, many pairs would exhibit
a different difference, and the attacker would not know with
certainty if the
difference is due to a different count, or due to the dynamic
noise layer.

\subsubsection{Equations Attack} 
\label{sec:eqAttack}

The equations attack from Section~\ref{sec:background} requires that the
amount of noise (the standard deviation of the noise) added to
each sum be well below a certain threshold.
We assume here that the equations attack also
requires that each query has enough conditions to select the specific
rows required by the attack: for each selected row, the query has
to specify the set of attribute values that uniquely
identify that row. Each such attribute value requires a condition,
and each condition adds additional noise to the sum.
Leaving out the details, in our experiments we found that
the equations attack fails because too much noise is added to each sum.

\subsubsection{Chaff Attacks}
\label{sec:chaff}

Section~\ref{sec:background} briefly described a chaff attack on the simple
sticky noise mechanism of Denning et.al.~\cite{denning1980secure}.
Because Denning's sticky noise is based on the entire query,
the attack only requires the addition, for instance, of a
condition that has no effect on the rows in a bucket, i.e.
\mytt{age<>1000}, \mytt{age<>1001} etc.
This specific simple chaff attack does not work with sticky layered
noise because the additional condition only creates an additional
noise layer, and doesn't affect the other noise layers.

The attack on Denning's sticky noise operates by changing the semantics of the
query conditions, and exploits the fact that the semantic change happens not
to effect the query result.
A chaff attack that simply changes the syntax of the query would
also work on Denning. For instance, the attacker could make a set
of queries each with 
syntactically different but semantically identical conditions like
\mytt{age=50}, \mytt{age+1=51}, \mytt{age+2=52} and so on.

Whether semantic or syntactic chaff attacks work on diffix-birch depends
entirely on what syntax is allowed in a query, and on
how noise layers are computed from that syntax. The diffix-birch design
in~\cite{diffix17} specified a very simple syntax, allowing only
conditions of the form \mytt{column operator constant}, where
the operators were limited to \mytt{=}, \mytt{<>}, \mytt{>},
\mytt{<}, \mytt{>=}, and \mytt{<=}. No math, string or datetime
functions, union semantics (\mytt{OR}), or JOIN's among other
things were allowed.
Those constraints prevented syntactic chaff attacks: the semantics
of a given query were always clear and sticky layered noise
defended against semantic chaff attacks (to the best of our
knowledge).

The design of diffix-birch in this paper allows a much richer subset
of SQL, and therefore a much larger attack vector for chaff
attacks. Chaff attacks are informally (and incompletely) addressed 
in other parts of this paper.

\subsection{Bucket Suppression}

Learning the strings or values that exist in a database is
very useful. Simply listing a column's values, however, violates
singling out when any of the values belong to a single user.
Therefore we wish to suppress any response where a value
represented by a single user.

A simple mechanism for this would be to have a threshold
for the number of distinct users that comprise a given
bucket below which the bucket is suppressed. We need to take
care, however, that the suppression or lack thereof doesn't
itself constitute a signal that might allow an analyst to
obtain individual user information. For instance,
a simple hard threshold, whereby all buckets with fewer than exactly
$K$ distinct users is suppressed, can leak the presence of
absence of a specific user
in the case where the analyst knows that there are either
$K$ or $K-1$ users with a given value. This $K$th user is
singled out in this case.

To avoid this, we use what we call a \emph{noisy threshold}.
This is a threshold whose value for any given bucket comes
from a Gaussian distribution around a mean threshold.
The noisy threshold used for bucket suppression, which we refer to
as the \emph{low-count threshold}, operates as follows.

If the number of distinct \mytt{uid}'s in the bucket is less than $T_h=2$,
then suppress the bucket. Otherwise, 
seed a sticky noise value $T_n$ with mean $\mu=4$ and standard deviation
$\sigma=0.5$ using the distinct \mytt{uid}'s from the
bucket:

\begin{footnotesize}
\begin{verbatim}
  seed = XOR(salt,hash(uid1), hash(uid2), ...)
\end{verbatim}
\end{footnotesize}

If the number of distinct \mytt{uid}'s in the bucket is less than
$T_n$, then suppress the bucket.

These specific values for $\mu$, $\sigma$, and $T_h$ are a matter
of policy. It is overall good to ensure that the mean $\mu$ is
three or four standard deviations from the hard threshold $T_h$
so that the hard threshold is rarely actually invoked and
therefore can't somehow be exploited. At the same time, it is safer
to have a larger $\sigma$, which then implies a larger mean. Unfortunately,
the larger the mean, the more suppression. Often columns have
a lot of distinct values that are populated by very few individuals,
and much of this information can disappear when $\mu$ is too large.
In addition, almost invariably analyst's like to look at small
groups, and so $\mu$ plays in important role in how fine-grained
the analysis can be. Therefore, we select relatively small values
for $\mu$, $\sigma$, and $T_h$.

\subsection{Baseline Noise and Noisy Threshold}
\label{sec:baseline}

The anonymizing aggregation functions use sticky layered noise
it two ways: to add noise to the final answer, and to select
groups of users using a noisy threshold. For both uses, diffix-birch
uses noise from the summed static and dynamic noise layers
described above. Specifically, diffix-birch produces a \emph{baseline
noise} value $N_b$ from which other noise values are derived
as needed.

The mean of the baseline noise $N_b$ is $\mu_{base}=0$, and its
standard deviation is $\sigma_{base}=1$.
If a different mean or standard deviation is required for a given
purpose, then the
baseline noise is simply multiplied by a factor to obtain
the required standard deviation, and the mean is adjusted accordingly.

For all of the noisy threshold uses in the anonymizing aggregation
functions, the hard threshold is set to $T_h=2$, and the noise
value is set to $N_b/2+4$ (mean of 4).

\subsection{Anonymizing Aggregation Functions}
\label{sec:anonFunc}

Diffix-birch implements the following anonymizing aggregation
functions: \mytt{count}, \mytt{sum}, \mytt{avg}, \mytt{stddev}, \mytt{min}, \mytt{max}, and \mytt{median}.
Each of these functions has a specific anonymization method that
trades off the desire to minimize distortion against the goal
of ensuring that the effect of any one individual in the
bucket cannot be easily detected.

The functions \mytt{count}, \mytt{sum}, \mytt{avg}, and \mytt{stddev}
are treated rather differently than the functions
\mytt{min}, \mytt{max}, and \mytt{median}. This is because the
former functions compute a value that is a composite of all values,
whereas the latter functions compute a value that belongs to only
a single user.

\subsubsection{count, sum, avg, and stddev}
\label{sec:sum}

The anonymization of these four functions are all based on 
that of \mytt{sum}. \mytt{count} is a \mytt{sum} of rows. \mytt{avg} is
\mytt{sum} divided by \mytt{count}. \mytt{stddev} is ultimately derived from
the \mytt{avg} function. Specifically for each row \mytt{stddev} computes
the square of the difference between each value
and the true average. It then computes the (anonymized)
\mytt{avg} of these differences, and reports the square root
of this \mytt{avg}.

There are two important but contradictory
considerations to make when computing
the anonymized \mytt{sum}. On the one hand, the amount of noise needs to be
proportional to the largest contribution of any user. 
This is to protect against the case where a difference
attack can isolate a user that contributes an unusually large amount
to the sum.

\begin{figure}
\center
\begin{footnotesize}
  \begin{tabular}{l | l}
\hline
SELECT sum(salary) &
SELECT sum(salary) \\

FROM table &
FROM table \\

WHERE dept = 'CS' AND &
WHERE dept = 'CS' \\

\hspace{1.0cm} warn = 1 AND &
\hspace{1.0cm} warn = 1 \\

\hspace{1.0cm} gender = 'M'  & \\
\hline
  \end{tabular}
\end{footnotesize}
\caption{Difference attack to learn whether the isolated woman has had a
disciplinary warning given that she has a high salary.}
\label{fig:diffsum}
\end{figure}

For instance, consider the attack of Figure~\ref{fig:diffsum} which
isolates the one woman in the CS department and tries to learn
if she has had a disciplinary warning. She is excluded from the
left query and included in the right if she has had a warning.
Suppose that the amount
of noise added is proportional not to the largest salary but to
the average salary. Suppose also that the woman's salary is
substantially higher than the average, say \$400K versus
\$125K. If the difference between the reported sums is around
\$400K or higher, then with very high probability the woman
is present in the right-hand query and therefore has had a
warning. This high probability comes from the fact that \$400K
is more than three standard deviations from \$125K, so the
probability that the difference is due purely to noise and
not the contribution of the woman's salary to the sum is
very small.

On the other hand, making the noise proportional to the largest
value causes a problem when that value is an outlier.
For instance, suppose instead that the woman's \$400K
salary in the attack
of Figure~\ref{fig:diffsum} is twice that of the
next highest salary, and that the analyst knows this fact.
Now suppose that the reported sum
of the right-hand query is \$800K greater than that of the
left-hand query. If the woman is present in the right-hand
query, then this difference is due to her salary being
included plus one standard deviation of noise. If the woman
is not present in the right-hand query, then the difference
would be due to four standard deviations of noise, a very
low probability. Therefore the analyst can conclude with
high probability that the woman has had a warning.

\begin{figure}[tp]
\begin{center}
\includegraphics[width=0.45\textwidth]{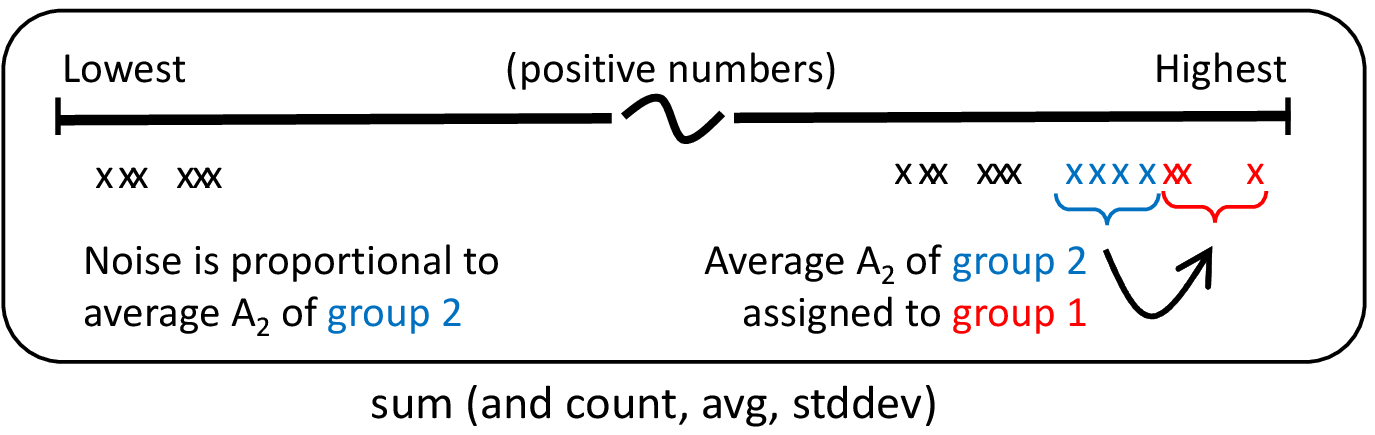}
\caption{To compute sum (and count, avg, and stddev which are based on sum), the high outliers (group 1) are "flattened" by assigning the value of the average $A_2$ of the next highest values (group 2). Noise is proportional to $A_2$.
}
\label{fig:sum}
\end{center}
\end{figure}

Therefore, to effectively hide any given user in a \mytt{sum},
it is necessary to first remove outliers, and then add
noise proportional to the remaining highest values.
Simply removing outliers, however, results in more distortion
than necessary. Instead, we \emph{flatten} the values
of a few distinct users with the highest values (group 1
in Figure~\ref{fig:sum} so that
they are comparable to those of the next few distinct users
with the highest values (group 2).

The \mytt{sum} (and \mytt{count}) function has two phases, a pre-processing
phase and a summing phase.  The input to the summing phase is:

\begin{enumerate}
\item A table
with two columns, a \emph{uid} column and a \emph{value} column.
The uid's are distinct, and the value contains each user's total
contribution to the count or sum.  
\item A \emph{baseline} sticky noise value $N_b$ derived from
the noise layers (see Figure~\ref{fig:baseflow} and
Section~\ref{sec:baseline}).
$N_b$ has a standard deviation of $\sigma = \sqrt L$, where $L$ is
the number of noise layers.
The actual noise is a multiplicative factor of $N_b$.
\end{enumerate}

Depending on the function,
the pre-processing phase produces the value as follows:

  \begin{description}
\item [\mytt{count(distinct uid)}:] All values are '1'. 
\item [\mytt{count(*)}:] Each value is the number of rows for the uid. 
\item [\mytt{count(column)}:] Same as \mytt{count(*)}, except rows with NULL values are not counted. 
\item [\mytt{count(distinct column)}:] Same as \mytt{count(column)}, except duplicate values are removed before counting each user's rows. 
\item [\mytt{sum(column)}:] Each value is the sum of the values for the uid. 
\item [\mytt{sum(distinct column)}:] Same as \mytt{sum(column)}, except duplicate values are removed before summing each user's values. 
  \end{description}

The summing phase consists of the following steps:

\begin{enumerate}
\item Generate two noisy thresholds $T_1$ and $T_2$
(see Section~\ref{sec:baseline}).
\item Label the $T_1$ distinct users with the highest values \emph{group 1}.
\item Label the $T_2$ distinct users with the next highest values \emph{group 2}.
\item Define $A_2$ as the average of the values from \emph{group 2}.
\item \label{1s1} Replace the \emph{group 1} values with $A_2$.
\item Define $A_{all}$ as the average of all values (after replacement).
\item \label{1s2} Compute a noise value $N = N_b*max(1/2{A_2},A_{all}$).
\item Sum the values, and add noise $N$.
\end{enumerate}

If there are both positive and negative values, then the above computations
are done separately for the positive and negative values, using lowest
values instead of highest.
The actual added noise is the sum of the two noise values.

As described above, both the \mytt{count(distinct column)} 
and \mytt{sum(distinct column)} functions remove duplicate
values during pre-processing.
Duplicates are removed in such a way as to maximize
the resulting number of distinct uid's.  This minimizes the contribution
of any given user, allowing diffix-birch to minimize the noise.

Note that in the case of \mytt{count(distinct uid)}, there is no
distortion due to flattening. All values are 1, so the
replacement in step~\ref{1s1} has no effect.

The reason in step~\ref{1s2}
for computing the noise $N$ as proportional the max of 
the average of all values $A_{all}$ or one half the average of
group 2 $A_2$ is to further lower the amount of noise in the case
where the highest values are substantially higher than the average
value. The premise here is that if the large values are
substantially higher than the average, then there is some spread
in the values contributed by different users, and therefore some
uncertainty on the part of the analyst as to how much any given
user is contributing to the sum. This uncertainty then reduces
the need for the uncertainty inherent in the noise. This is
admittedly at this time an unfounded premise, and more
experimentation is needed to ensure that the premise is correct. 

\subsubsection{min, max, and median}

\begin{figure}[tp]
\begin{center}
\includegraphics[width=0.45\textwidth]{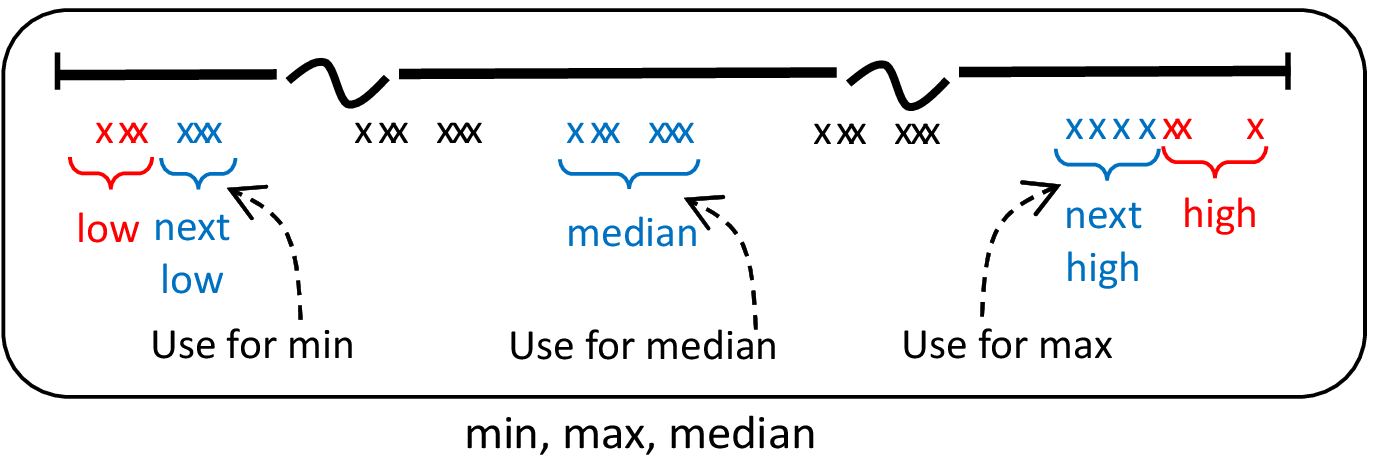}
\caption{To compute \mytt{min}, \mytt{max}, and \mytt{median}, the highest and lowest values are removed.  \mytt{min}, \mytt{max}, and \mytt{median} are then computed from the average of the appropriate group of values as shown.
}
\label{fig:minmaxmedian}
\end{center}
\end{figure}

Since \mytt{min} and \mytt{max} in principle could report the
value of an individual user, it is particularly important that
the effect of extreme values are hidden.
Diffix-birch removes noisy threshold numbers of distinct
users from both the high and low end of the value range
(see Figure~\ref{fig:minmaxmedian}).
The average value of the
next highest (lowest) noisy threshold number of distinct users
are then used to compute \mytt{max} (\mytt{min}).
Similarly, the \mytt{median} is computed from the average of the
values of a noisy threshold number of distinct users above and
below the true median.

Once the highest and lowest value are removed,
\mytt{max} is computed as follows:

\begin{enumerate}
\item Select a group of values
from the values of a noisy threshold number of distinct (next) highest users.
\item Compute the average $A$ of the group.
\item Compute the standard deviation $\sigma$ of the group.
\item \label{2s1} The anonymized max value is computed as $A$ with added noise ${(N_b * {\sigma})}/8$.
\end{enumerate}

The computation of the noise amount in step~\ref{2s1} needs some explanation.
$N_b$ is the noise value generated from the baseline noise.
If all the values in the group are the same, then the standard deviation
of the group is zero, and there is no noise added. In this case, the exact
value of the group is reported. This is safe, because the group represents
multiple distinct users and so any one user is hidden in the crowd.
Furthermore, the analyst does not know how many users constitute the group.
Indeed, if all the values in the averaged group and the removed group
are identical, then
\mytt{max} or \mytt{min} will report the correct true max or min.

If however, the values of the group are different, then we add a little noise
to help ensure that the analyst cannot somehow reverse engineer the specific
values in the group from the exact average $A_2$. In fact, we don't have
evidence that this reverse engineering would be possible, but neither
do we have evidence that it is not. So the noise is a fudge factor.
With experimentation, we may find that it is not needed, or for that
matter that it should be larger.

\subsubsection{median}

Like \mytt{min} and \mytt{max}, \mytt{median} can report the value
of a single user, and can report the correct true median. 
Also like \mytt{min} and \mytt{max}, the values for the highest and lowest
distinct users are first removed for computing \mytt{median} (see
Figure~\ref{fig:minmaxmedian}).
After that removal, the algorithm is as follows:

\begin{enumerate}
\item Order the rows from highest to lowest values.
\item Generate a noisy threshold $T$
(see Section~\ref{sec:baseline}).
\item Select the true median row and label it.
\item Label the rows for the $T$ distinct users above the true median, and the $T$ distinct users below the true median (the same user may appear once above and once below).
\item Compute $A$ as the average of the labeled rows.
\item Compute $\sigma$ as the standard deviation of the labeled rows.
\item The anonymized median value is computed as $A$ with added noise ${(N_b * \sigma)}/8$.
\end{enumerate}

\subsection{An Example}

\begin{figure*}[tp]
\begin{center}
\includegraphics[width=0.9\textwidth]{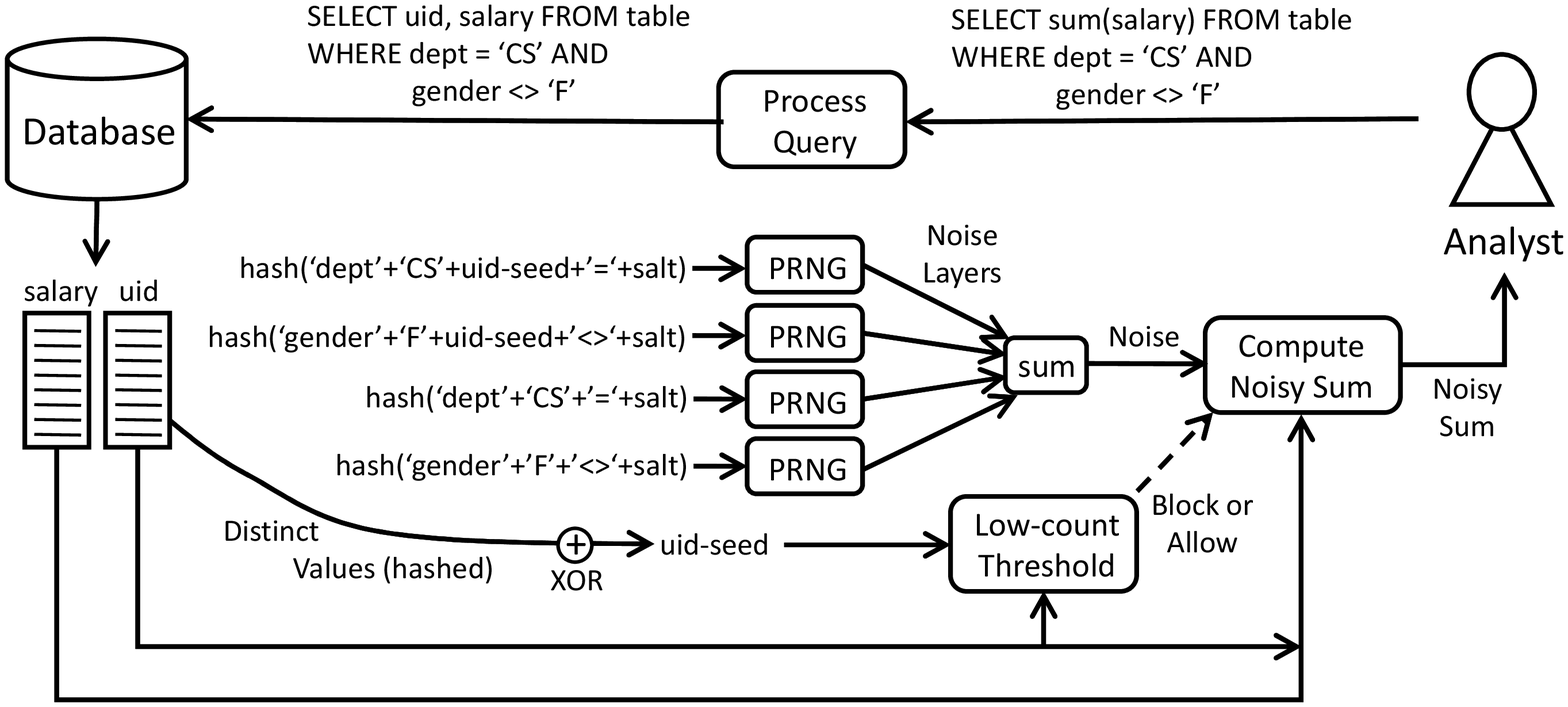}
\caption{Diffix-birch workflow for an example query}
\label{fig:noise}
\end{center}
\end{figure*}

Figure~\ref{fig:noise} illustrates an example query in diffix-birch.
The analyst requests the sum of salaries with two conditions
(on columns \mytt{dept} and \mytt{gender}).

Diffix-birch processes the query, both to ensure that it conforms to the
SQL restrictions placed by diffix-birch, and to modify the query so that
the data required for anonymization is retrieved from the database.
Both the low-count threshold and the anonymized \mytt{sum} function require
the list of \mytt{uid}'s, and the \mytt{sum} function additional
requires the list of salary values. As a result, the query is
modified so as to request this data from the database. (Note that 
this is somewhat a simplification of what happens in our implementation.)

The contents of the
\mytt{uid} column are used to generate the \mytt{uid-seed} for the low-count
threshold check as well as for the dynamic noise layers. The low-count
threshold check determines whether the bucket needs to be suppressed.

In this particular example, the seeds for the layered noise can be
generated from examining the SQL itself. Two static and dynamic
noise layers are generated, one each for the two conditions.

\section{Details}
\label{sec:details}

The previous section describes the basic concepts of diffix-birch,
giving examples for queries with simple conditions: \mytt{AND}'d
conditions in a \mytt{WHERE} clause with no math or other functions.
With the exception of a few updates,
the concepts and examples essentially cover the material described
in~\cite{diffix17}. This section describes how diffix-birch is applied to
a much richer SQL syntax. 

The complete diffix-birch syntax is specified in Appendix~\ref{sec:syntax}.

There are a number of restrictions on how the syntax specified
in Appendix~\ref{sec:syntax}
may be used. As a rule, the purpose of these restrictions
is to prevent an attacker from either generating conditions that
diffix-birch does not recognize as conditions, or generating conditions
whose \emph{semantics} diffix-birch mis-interprets.

An example of the latter is given in Section~\ref{sec:chaff},
where for instance the conditions \mytt{WHERE age=20},
\mytt{WHERE age+1=21}, \mytt{WHERE age+2=22}
etc. are all semantically identical and so must produce the same
noise layer.

An example of the former is the query shown in Figure~\ref{fig:badMath},
where the two queries both provide row counts of users age 30 or
40, but where the upper query has no explicit condition at all.
We refer to attacks that emulate conditions without diffix-birch
realizing it as \emph{backdoor attacks}.

These restrictions are described throughout this section.

\subsection{Terminology}

A \emph{positive} condition is one that includes rows that match
the condition. For instance, 

\begin{footnotesize}
\begin{verbatim}
WHERE column = value
WHERE column IN (values)
WHERE column LIKE value
\end{verbatim}
\end{footnotesize}

A \emph{negative} condition is one that excludes rows that match
the condition. For instance, 

\begin{footnotesize}
\begin{verbatim}
WHERE column <> value
WHERE column NOT IN (values)
WHERE column NOT LIKE value
\end{verbatim}
\end{footnotesize}

An \emph{equality} is a condition that requires an exact match to
include or exclude rows. Diffix-birch
has equality operators \mytt{=} and \mytt{$<>$}.

An \emph{inequality} is a condition with operators like less than
or greater than. Diffix-birch places a restriction on inequalities in that
the inequalities must express a \emph{range}: both lower and upper
inequality boundaries must be specified. Thus the condition
\mytt{WHERE age BETWEEN 10 AND 20} is allowed, but the condition
\mytt{WHERE age $<$ 20} is not allowed. Both positive and negative
ranges may be expressed (i.e. \mytt{BETWEEN} and \mytt{NOT BETWEEN}).

A query condition may be \emph{clear} or \emph{unclear}. Simply
put, a clear condition is one where diffix-birch understands the intent
of the condition and can therefore derive the condition's noise layers
directly by examining the SQL. What passes as clear therefore
depends on the sophistication of the diffix-birch SQL compiler (see
Section~\ref{sec:clear} below).

\subsection{Current scope of clear queries}
\label{sec:clear}

As of this writing, the diffix-birch SQL compiler is not very sophisticated.
To be considered clear, the column in the condition must either be the
native column, or a native column being operated on by
one of the following string operations:
\mytt{right}, \mytt{left}, \mytt{ltrim}, \mytt{rtrim}, \mytt{btrim},
\mytt{trim}, \mytt{substring}, \mytt{upper}, or \mytt{lower}.
To be considered clear, the condition must also have a simple
constant (not operated on by math or any function).
The upper four queries of Figure~\ref{fig:clear} are clear,
while the bottom two are not.

\begin{figure}
\center
\begin{footnotesize}
  \begin{tabular}{l | l}
\hline
SELECT sum(salary) &
SELECT count(*) \\
FROM table &
FROM table \\
WHERE left(date,4) = '2009' &
WHERE age = 64 \\
\hline
SELECT avg(salary) &
SELECT gender, count(*) \\
FROM table &
FROM table \\
WHERE age &
WHERE age \\
\hspace{0.3cm} NOT IN (25,26,27) &
\hspace{0.3cm} BETWEEN 10 AND 20 \\
\hline
\hline
SELECT sum(salary) &
SELECT count(*) \\
FROM table &
FROM table \\
WHERE left(date,4) $<>$ year &
WHERE sqrt(age) = 8 \\
\hline
  \end{tabular}
\end{footnotesize}
\caption{The conditions for the top four queries are clear.
The conditions for the bottom two are unclear.
}
\label{fig:clear}
\end{figure}

The reason we specifically allow the string functions is because
in our experience it is quite common for analysts to select
sub-strings from a string. This is because strings frequently
have an internal structure that the analyst wants to deconstruct
(for instance a date encoded as a string, as in the example
of Figure~\ref{fig:clear}). In so doing we are conceptually treating the
substring as a separate column.
Nevertheless, when there is one of these string functions present,
diffix-birch adds an extra noise layer to defend against
potential attacks that try to exploit the function
(Sections~\ref{sec:stringFuncs} and \ref{sec:upperLower}).

\begin{figure}
\center
\begin{footnotesize}
  \begin{tabular}{l}

\hline
SELECT count(*), age\_30\_or\_40 FROM ( \\
\hspace{0.3cm}  SELECT uid, \\
\hspace{0.3cm}  (age\_30 + age\_40) \% 2 as age\_30\_or\_40 FROM ( \\
\hspace{0.6cm}    SELECT uid, \\
\hspace{0.6cm}    floor((age\_greater\_29 + age\_less\_31) / 2) AS age\_30, \\
\hspace{0.6cm}    floor((age\_greater\_39 + age\_less\_41) / 2) AS age\_40 \\
\hspace{0.6cm}    FROM ( \\
\hspace{0.9cm}      SELECT \\
\hspace{0.9cm}      uid, \\
\hspace{0.9cm}      ceil((age - 29) / 100) AS age\_greater\_29, \\
\hspace{0.9cm}      ceil(0 - (age - 31) / 100) AS age\_less\_31, \\
\hspace{0.9cm}      ceil((age - 39) / 100) AS age\_greater\_39, \\
\hspace{0.9cm}      ceil(0 - (age - 41) / 100) AS age\_less\_41 \\
\hspace{0.9cm}      FROM table \\
\hspace{0.6cm}    ) x \\
\hspace{0.3cm}  ) y \\
) z \\
GROUP BY age\_30\_or\_40; \\
\hline
SELECT count(*) \\
FROM table \\
WHERE age = 30 OR age = 40 \\
\hline
  \end{tabular}
\end{footnotesize}
\caption{Backdoor Conditions: The upper and lower queries both produce counts of individuals aged 30 or 40. The upper query is disallowed by limiting the number of discontinuous functions combined with math and constants that may appear in a query.
}
\label{fig:badMath}
\end{figure}

Diffix-birch supports use of the \mytt{HAVING} clause in sub-queries.
A \mytt{HAVING} clause necessarily implies an aggregation function
and associated \mytt{GROUP BY} which
must operate on the \mytt{uid}. Queries
where this is not the case are rejected. 
The rules for when a \mytt{HAVING} clause is clear in the current
implementation are the same as for the \mytt{WHERE} clause, with
the exception that an aggregation function is allowed, and the
constraint that the aggregation function must operate on a raw
column value. For example, neither query of
Figure~\ref{fig:havingClear} is considered clear. (Both would be
clear if the \mytt{+20} was removed.)

\begin{figure}
\center
\begin{footnotesize}
  \begin{tabular}{l | l}
\hline

SELECT mxage, count(*) &
SELECT mxage, count(*) \\
FROM ( &
FROM ( \\
\hspace{0.3cm}  SELECT uid, &
\hspace{0.3cm}  SELECT uid,  \\
\hspace{0.6cm}    max(age) AS mxage &
\hspace{0.6cm}    max(age+20) AS mxage \\
\hspace{0.3cm}  FROM table &
\hspace{0.3cm}  FROM table \\
\hspace{0.3cm}  GROUP BY uid  &
\hspace{0.3cm}  GROUP BY uid  \\
\hspace{0.3cm}  HAVING mxage+20 = 65)t &
\hspace{0.3cm}  HAVING mxage = 65)t \\
GROUP BY mxage &
GROUP BY mxage \\

\hline
  \end{tabular}
\end{footnotesize}
\caption{Neither of these queries are clear, because of the math.
Both would be clear if the \mytt{+20} were removed.
}
\label{fig:havingClear}
\end{figure}

\subsection{AND'd conditions}

In diffix-birch, the \mytt{OR}
operation is not allowed. There is no fundamental reason
for this. Indeed the \mytt{IN} operation, which has union
semantics, is allowed.

A diffix-birch query consists of the intersection of
a set of conditions, where the condition operators may
be \mytt{=}, \mytt{BETWEEN} (or the equivalent pair of inequalities),
\mytt{IN}, \mytt{LIKE}, \mytt{IS NULL}, and the corresponding
negative variants (\mytt{NOT IN} etc.).

A diffix-birch query may have zero or more conditions.
If a query has zero conditions, then a single dynamic
noise layer is used. The \emph{seed components} in this
case consist of the table name, the hashed and XOR'd
\mytt{uid}'s, and the secret salt.

When columns are selected, they are treated as positive AND'd
conditions for the purpose of seeding noise layers.
For instance, the left-hand query from Figure~\ref{fig:select}
generates a set of buckets, one for each unique pair of
\mytt{age} and \mytt{salary}. For each pair of values, noise layers
are seeded as though those values appeared in a \mytt{WHERE} clause
(i.e. the query on the right).

\begin{figure}
\center
\begin{footnotesize}
  \begin{tabular}{l | l}
\hline
SELECT age, salary, count(*) &
SELECT count(*) \\

FROM table &
FROM table \\

GROUP BY age, salary & 
WHERE age = 20 AND \\

 & \hspace{1.0cm} salary = 100000 \\

\hline
  \end{tabular}
\end{footnotesize}
\caption{Selected columns are implicitly AND'd positive conditions, and are treated as such for the purpose of seeding noise layers. The bucket produced by the left-hand query where \mytt{age=20} and \mytt{salary=100000} is seeded identically to that of the right-hand query.
}
\label{fig:select}
\end{figure}

Note that selected columns do not require an associated
anonymizing aggregation function. For instance, the query
on the left-hand side of Figure~\ref{fig:noagg} is valid.
In this case, however, for the purpose of anonymization
the query is treated as though the \mytt{count(*)}
anonymizing aggregation function were present. The only
difference is in the layout of the query answer. For example,
the answer to the right-hand query of Figure~\ref{fig:noagg}
would have a single row for each age. If the noisy count
for \mytt{age=20} from the right-hand query were 86, then
the left-hand query would simply produce 86 rows each with
\mytt{age=20}.

\begin{figure}
\center
\begin{footnotesize}
  \begin{tabular}{l | l}
\hline
SELECT age &
SELECT age, count(*) \\

FROM table &
FROM table \\

 & GROUP BY age \\

\hline
  \end{tabular}
\end{footnotesize}
\caption{When no anonymizing aggregation function is present (left-hand
query), then for the purpose of anonymization the query is treated as though \mytt{count(*)} was used (right-hand query).
}
\label{fig:noagg}
\end{figure}

\subsection{Floating versus SQL Inspection}

In Figure~\ref{fig:noise}, diffix-birch is shown as requesting the \mytt{uid}
column from the database in order to compute the \mytt{uid}
seed component of the dynamic noise layers. We refer to this mechanism of
requesting a column as \emph{floating} the column. Every query
requires that at least the \mytt{uid} column is floated.
In the example of Figure~\ref{fig:noise}, however, the other seed
components are generated simply by inspecting the SQL itself.
This is possible because the conditions in the query are clear.

There are cases, however, where the seed components cannot be derived
from SQL inspection alone. One such case is that of selected columns
being treated as implicit positive AND'd conditions as shown in
Figure~\ref{fig:select}. The selected columns are floated and the values
returned by the database are inspected and used as seed components.

Another example is where conditions are unclear. For instance, in
the left-hand query of Figure~\ref{fig:float}, the condition
\mytt{WHERE age+1=26} is unclear because of the math. In this
case, the \mytt{age} column is floated (along with the \mytt{uid}
column) in order to compute the seed components of the noise layers
for the \mytt{age} condition (right-hand query). In this case, the
floated value for \mytt{age} used in the seed would be 25, which
indeed captures the "intent" of the condition.
Floating gives the analyst access to a variety
of math, string, and datetime operations while preventing chaff
attacks that exploit those operations.

 \begin{figure}
 \center
 \begin{footnotesize}
   \begin{tabular}{l | l}
\hline
 SELECT count(*) &
 SELECT uid, age \\
 
 FROM table &
 FROM table \\
 
 WHERE age+1 = 26 &
 WHERE age+1 = 26 \\
\hline
   \end{tabular}
 \end{footnotesize}
 \caption{Because the condition is unclear, diffix-birch extracts and examines (floats) the column contents to derive the \mytt{value} seed component. When the analyst submits the left-hand query, diffix-birch transforms it to the right-hand query to extract the \mytt{age} column (and \mytt{uid} column) from the database.
}
 \label{fig:float}
 \end{figure}

In the case of a sub-query with a \mytt{GROUP BY}, the raw column
cannot be floated because it no longer exists after the aggregation
function.  For instance in the left-hand query of
Figure~\ref{fig:aggFloat}, the \mytt{avg} function in the sub-query transforms
the individual rows into aggregate rows. It is therefore no longer possible
to float the individual rows and use their values for seeding.

 \begin{figure}
 \center
 \begin{footnotesize}
   \begin{tabular}{l | l}
\hline
 SELECT avtime, count(*) &
 SELECT uid, avtime, mn, mx, ct \\
 FROM ( &
 FROM ( \\
\hspace{0.3cm} SELECT uid, &
\hspace{0.3cm} SELECT uid, \\
\hspace{0.3cm} avg(time) AS avtime &
\hspace{0.3cm} avg(time) AS avtime, \\

\hspace{0.3cm} FROM taxi\_rides &
\hspace{0.3cm} min(time) AS mn, \\

\hspace{0.3cm} GROUP BY uid) t &
\hspace{0.3cm} max(time) AS mx, \\

GROUP BY avtime &
\hspace{0.3cm} count(time) AS ct \\

 & \hspace{0.3cm} FROM taxi\_rides \\
 & \hspace{0.3cm} GROUP BY uid) t \\

\hline
   \end{tabular}
 \end{footnotesize}
 \caption{The \mytt{avg} aggregation function in the analyst's query (left side query) prevents floating the individual row values for the column \mytt{time}. Instead, \mytt{min}, \mytt{max}, and \mytt{count} aggregates are floated and used to seed noise layers (right side query).
}
 \label{fig:aggFloat}
 \end{figure}
 
One way to deal with this would be to somehow incorporate the aggregation
function into the seed. For instance, for the left-hand query of
Figure~\ref{fig:aggFloat}, the noise layer could additionally be seeded
by the name of the aggregation function. The problem with this is that
several aggregation functions can produce the same results. In any column
where there is one row per distinct user, then \mytt{min}, \mytt{max}, \mytt{avg},
and \mytt{median} for instance all produce the same value. This allows the
analyst to get four noise samples and so is a chaff attack, albeit limited
by the number of different aggregation functions.

Rather, diffix-birch floats pre-specified aggregates, specifically
\mytt{min}, \mytt{max}, and \mytt{count}, and uses the associated
values in the seed. Pre-specified
aggregates are chosen so that the analyst cannot influence what gets
floated. These three particular aggregates are chosen because they work with
numeric, text, and datetime column types, and at the same time provide good
"coverage" of the column contents. 

The specific attack on \mytt{HAVING} that we want to avoid is
one where the analyst isolates a user with two queries in such a way
that the seeds for the noise layers associated with the aggregated
column are the same. This could happen if the only difference between
the two queries is the choice of aggregation function, the two
aggregation functions change only a single row (that of the victim),
but at the same time does not change the \mytt{min}, \mytt{max}, or \mytt{count}
for the victim. While theoretically possible, we take it as a given
that the probability of this occurring (and the analyst knowing it
occurs) is negligible and does not need to be defended against.

\subsection{Positive equality}
\label{sec:posEq}

Positive equalities are allowed. 
Positive equalities do not need to be clear.
The seed components for the static
noise layer, not including the salt, are:

\begin{footnotesize}
\begin{verbatim}
   [table_name, column_name, value, value, 1]
\end{verbatim}
\end{footnotesize}

If the condition is clear, then the value
is simply extracted from the SQL and used to seed the
noise layer.
The same seed components are used whether the condition is
in a \mytt{WHERE} or a \mytt{HAVING} clause.
If the condition is not clear, then the column, or in the
case of \mytt{HAVING} its \mytt{min/max/count} aggregates,
is floated and the value is taken from the database.

The reason that the value is repeated, and the 1 included, can
be explained as follows. Suppose that only a single instance
of the value was used (i.e. \mytt{[value]} instead of
\mytt{[value,value,1]} as is the case in the simplified
seed description from Section~\ref{sec:stick}, where \mytt{value='CS'}).
Suppose further
than the table has one row per user. In that case, the
analyst could compose a query containing an unclear \mytt{HAVING} that
forces diffix-birch to float \mytt{max/max/count} aggregate values rather than
the native column values. This would produce a seed that is
different from that produced with the corresponding \mytt{WHERE} clause,
thus giving the analyst another random sample. By using
\mytt{[value,value,1]} instead of \mytt{[value]}, the same seed results whether
the analyst uses an unclear \mytt{HAVING} or a \mytt{WHERE} clause.

The seed for the dynamic layer of course additionally contains the
XOR'd hashes of the \mytt{uid}'s. For brevity,
in what follows unless otherwise
specified we exclude the secret salt, table name, and
column name from the specification of the static seed components,
and additionally exclude the XOR'd hashes of \mytt{uid}'s from
the specification of the dynamic seed components.

\subsection{Negative equality}
\label{sec:negEq}

Negative equalities are allowed. Negative equalities must always
be clear (in our current implementation).
Examples of queries with negative equalities are the following:

\vspace{0.3cm}
\begin{footnotesize}
  \begin{tabular}{l | l}
SELECT count(*) &
SELECT sum(salary) \\
FROM table &
FROM table \\
WHERE age $<>$ 64 &
WHERE left(date,4) $<>$ '2011' \\
  \end{tabular}
\end{footnotesize}
\vspace{0.3cm}

The difficulty with generating seeds for negative conditions in
general is that floating the column from the query as-is does not
tell diffix-birch what the condition is excluding, but rather what happens
to be included. In other words, SQL inspection of
the condition \mytt{WHERE age$<>$64} tells diffix-birch what the condition
excludes, while floating of the \mytt{age} column tells diffix-birch what
is included as a result of the negative condition.

To use floating (and therefore allow unclear conditions), diffix-birch
would have to \emph{reverse the condition} (change it from negative
to positive), and \emph{probe} the database with the modified
query to determine what the condition is meant to exclude.
For instance, the unclear condition on the left query below would
be changed to the unclear condition on the right. An initial
probe query would be made floating the \mytt{age} column. The value
returned by the database would be \mytt{65}, and this could then be
used to seed the noise layer.

\vspace{0.3cm}
\begin{footnotesize}
  \begin{tabular}{l | l}
SELECT count(*) &
SELECT count(*) \\
FROM table &
FROM table \\
WHERE age+5 $<>$ 69 &
WHERE age+5 = 69 \\
  \end{tabular}
\end{footnotesize}
\vspace{0.3cm}

The decision to not implement a probing approach is simply one of
prioritization. There is extra overhead (the additional query) and
extra complexity. Given that in the future we can expand the
set of queries that are clear with a smarter SQL compiler, the
question of whether it is worth implementing a probing approach
is still uncertain.

The seed components for the static noise layer include:

\begin{footnotesize}
\begin{verbatim}
   [value]
\end{verbatim}
\end{footnotesize}

Note that here there is no need to replicate the \mytt{value} and
include \mytt{1}. This is because there is no floating, and so never
a need to float \mytt{min/max/count}.

\subsection{Positive range}
\label{sec:posRange}

Inequalities must always be expressed as ranges. So the condition
\mytt{WHERE age BETWEEN 10 and 20} is allowed (or the equivalent using
\mytt{$>=$} and \mytt{$<=$}), but \mytt{WHERE age $<=$ 20} alone (without
a corresponding \mytt{$>=$}) is not allowed. The reason for this
restriction is because of \emph{snapped alignment}. Inequalities
in diffix-birch are forced into a pre-determined (snapped) set of exponentially
growing range sizes and offsets~\cite{diffix17}. Numbers fall on
intuitive boundaries like 0.1, 0.2, 0.5, 1, 2, 5, 10, 20, 50 etc.,
and datetimes fall on natural boundaries (second, hour, day etc.)
and intuitive sub-boundaries. Simply put, in order to for a
snapped size to be enforced, both edges of a range must be
specified by the analyst.

Ranges must be clear.  The seed components for the static
noise layer include:

\begin{footnotesize}
\begin{verbatim}
[lower_value, upper_value]
\end{verbatim}
\end{footnotesize}

Note that if \mytt{$>=$}, \mytt{$<=$} syntax is used, then two conditions
are treated as one.

The reason for snapped alignment is to prevent chaff attacks
whereby the analyst repeats a series of queries, each with a
slightly larger range, but in such a way that the small range change
does not change the set of rows filtered by the condition. As
a simple example, if the analyst knows that a given numerical
column contains integers, the analyst could increase the range
by increments of 0.01, thus obtaining many noise samples to
average out.

As of this writing, we have no good ideas on how to allow unclear
range queries. Unlike equalities, a range naturally allows 
a variety of different values to pass (those that fall in the range).
If diffix-birch could deduce from floating the range's column what the
intended range is, then diffix-birch could 1) ensure that the range is
snapped, and 2) use the range to seed the associated noise layers.
The problem is that it is hard to deduce the intended range.
For instance, suppose that diffix-birch observes values between 2
and 9 from floating the column. Diffix-birch cannot tell if the
specified unclear range in the query is 0-10 and therefore
properly snapped, or 2-9 and therefore not snapped, or even 0-20
because the range is properly snapped but just happens to have
no values between 10 and 20.

\subsection{Negative range}

Negative ranges (i.e. \mytt{NOT BETWEEN}) are allowed. Negative
ranges must of course also be clear. The seed components
are identical to those for positive ranges, but with the
addition of a symbol denoting that the range is negative:

\begin{footnotesize}
\begin{verbatim}
[lower_value, upper_value, ':<>']
\end{verbatim}
\end{footnotesize}

\subsection{IN clause}
\label{sec:in}

The \mytt{IN} clause is equivalent to a series of \mytt{OR}'d positive
equalities. For example, the following two queries are identical:

\vspace{0.3cm}
\begin{footnotesize}
  \begin{tabular}{l | l}
SELECT count(*) &
SELECT count(*) \\
FROM table &
FROM table \\
WHERE gen = 'M' AND &
WHERE gen = 'M' AND \\
\hspace{1.0cm} age IN (30,31) &
\hspace{1.0cm} (age = 30 OR age = 31)\\
  \end{tabular}
\end{footnotesize}
\vspace{0.3cm}

The \mytt{IN} clause must be clear.

There are two ways in which an analyst could attempt a difference
attack based on the \mytt{IN} clause. One is to remove the entire
clause. Another is to remove or change one or more
of the \emph{elements} within the \mytt{IN}. As such, there are
noise layers associated with the entire clause (the
\emph{per-clause} noise layers), and a noise layer associated with
each element (the \emph{per-element} noise layers).

Regarding the per-clause layer, diffix-birch creates a static
noise layer (no dynamic layer) by floating the associated column.
The noise layers are seeded with the following components:

\begin{footnotesize}
\begin{verbatim}
[val1, val2, ...]
\end{verbatim}
\end{footnotesize}

Where \mytt{val1}, \mytt{val2} etc. are the floated column values.
Note that
if the \mytt{IN} clause has only a single element then it is treated
identically to the corresponding positive equality
(Section~\ref{sec:posEq}).

The reason that the \mytt{IN} column is floated, even though it is
clear, is to defend against an averaging attack using chaff elements.
For instance, if diffix-birch used SQL inspection to compose the
per-clause noise layer, the analyst could submit a series of
queries with \mytt{age IN (50,1000)}, \mytt{age IN (50,1001)}, etc.

Regarding the per-element noise layers, diffix-birch generates a dynamic
noise layer for each element (no static layer). The seed
components include:

\begin{footnotesize}
\begin{verbatim}
[value]
\end{verbatim}
\end{footnotesize}

where \mytt{value} is the element itself.

The reason we don't require a per-clause dynamic noise layer is because
any one of the per-element dynamic noise layers is effective in
defending against a difference attack based on removing the clause.

The reason we don't require a per-element static noise layer is because
a set of per-element static noise layers, when summed together, behaves
like a per-group static noise layer, but with more noise (a higher
standard deviation). By having a single per-clause static noise layer,
we reduce the total amount of noise.

\subsection{NOT IN clause}
\label{sec:notin}

A \mytt{NOT IN} clause is equivalent to a set of \mytt{AND}'d negative
equalities. For example, the following two queries are equivalent.

\vspace{0.3cm}
\begin{footnotesize}
  \begin{tabular}{l | l}
SELECT count(*) &
SELECT count(*) \\
FROM table &
FROM table \\
WHERE age NOT IN (30,31) &
WHERE age $<>$ 30 AND \\
 & \hspace{1.0cm} age $<>$ 31 \\
  \end{tabular}
\end{footnotesize}
\vspace{0.3cm}

Recognizing this, diffix-birch treats any \mytt{NOT IN} clause as its
equivalent set of \mytt{AND}'d negative equality conditions
(Section~\ref{sec:negEq}).

\subsection{LIKE clause}
\label{sec:like}

The \mytt{LIKE} and \mytt{ILIKE} clauses enable a
difference attack whereby a victim
is isolated by exploiting a unique string associated with the victim.
For example, suppose that the database has a column \mytt{name}
consisting of last names.
Suppose that the following conditions hold:

\begin{enumerate}
\item The analyst has prior knowledge of the \mytt{name} column,
\item the column contains a number of Murry's (enough
to avoid low-count filtering), 
\item the column contains a single McMurry,
\item the column contains no other last names that end with Murry.
\end{enumerate}

Under these conditions, the analyst could execute the difference
attack on the victim McMurry with the queries shown in
Figure~\ref{fig:like}. The left-hand query excludes McMurry
while the right-hand query includes him or her because of the
'\%' wildcard symbol which matches on any zero or more of any
character.
As a result, whichever bucket matches McMurry's gender will
contain McMurry.
\mytt{LIKE} also permits the '\_' wildcard, which matches on exactly
one of any character.

\begin{figure}
\center
\begin{footnotesize}
  \begin{tabular}{l | l}
\hline

SELECT gender, count(*) &
SELECT gender, count(*) \\
FROM table &
FROM table \\
WHERE name LIKE 'Murry' &
WHERE name LIKE '\%Murry' \\
GROUP BY gender &
GROUP BY gender \\

\hline
  \end{tabular}
\end{footnotesize}
\caption{In this difference attack, the \mytt{LIKE} comparator isolates the
single McMurry in the case where there are multiple Murry's.
}
\label{fig:like}
\end{figure}

Note that, at least with last names, the conditions for this attack
exist surprisingly often. For instance, in a database of 250K last
names with a distribution published by the US census, 0.2\% of the
names had the attack condition. Similar results were measured for
Twitter hashtags.

Diffix-birch defends against this form of difference attack by more-or-less
creating a noise layer per wildcard in the \mytt{LIKE} expression.
The reason we do it this way, rather than try to create a single
noise layer from the complete set of wildcards, is that it is too
easy for an analyst to create chaff wildcard expressions otherwise.
For instance, the expressions 'Mur\_y', 'Murr\_', 'Mu\%rry' etc.
might all filter the same rows. If we had a single noise layer per
expression each seeded differently, then the noise could be averaged out.

In creating the noise layers, we want to try to best capture the
role that each wildcard plays in the expression. Note in particular
that the '\%' wildcard can often be inserted anywhere without having
any effect on the filtered rows. For instance, suppose that the
analyst wants to do the condition \mytt{col LIKE 'abc\_de'}, but would
like to average away the noise layer associated with the '\_' wildcard
using chaff. The analyst could try '\%abc\_de', '\%a\%bc\_de',
'\%ab\%\%c\_de', etc. It could well be that none of the '\%' wildcards
have any effect, but each new combination for instance pushes the
'\_' wildcard into a different string index position. Therefore we want
to ensure that this kind of chaff attack also doesn't work.

Our solution is to use the index position of the wildcard symbol to
seed its associated noise layer, but to ignore '\%' symbols when
determining the index position.  We also automatically ignore any
repeated '\%' symbols, since such symbols have no effect on the operation
of the \mytt{LIKE}. The creation of the noise layers then takes the
following steps:

\begin{enumerate}
\item Modify the original expression by:
  \begin{enumerate}
  \item Removing all repeating '\%' symbols.
  \item Modifying any sequence of characters containing both '\%' and '\_' symbols to contain instead a single '\%' at the beginning of the sequence, followed by the same number of '\_' symbols.
  \end{enumerate}
\item Generate a temporary expression by removing all '\%' symbols, and:
  \begin{enumerate}
  \item Compute $N$ as the number of characters in the temporary expression.
  \item Compute the index position $i$ of each character in
        the temporary expression.
  \end{enumerate}
\item For each '\%' symbol in the modified original expression,
      seed a noise layer using the following:
  \begin{enumerate}
  \item $N$.
  \item The position $i$ of the character preceding the '\%' symbol.
  \item The symbol '\%'.
  \end{enumerate}
\item For each '\_' symbol, seed a noise layer using the following:
  \begin{enumerate}
  \item $N$.
  \item The position $i$ of the '\_' symbol.
  \item The symbol '\_'.
  \end{enumerate}
\end{enumerate}

As a result of this algorithm, the three expressions
'\%abc\_de', '\%a\%bc\_de', and '\%ab\%\%c\_de' all compute the
same noise layer for the '\_' symbol because the index position
is always derived from the base string 'abc\_de'. The apparnetly
unavoidable downside of this approach is that it can create
a lot of noise layers. Analysts must be aware of this and strive
not to include unnecessary wildcards in \mytt{LIKE} expressions.

The \mytt{LIKE} clause must be clear.

\subsection{NOT LIKE clause}

The \mytt{NOT LIKE} and \mytt{NOT ILIKE} clauses
must also be clear. \mytt{NOT LIKE} is seeded
identically to \mytt{LIKE}, with the exception that the \mytt{:not} symbol is
included as a seed component.

\subsection{Character removing string functions}
\label{sec:stringFuncs}

The current implementation of diffix-birch supports a number of string
functions that can remove characters from the string. They include
\mytt{right}, \mytt{left}, \mytt{ltrim}, \mytt{rtrim}, \mytt{btrim}, \mytt{trim}, and \mytt{substring}.
All of these functions may appear in positive
and negative conditions, including those that are otherwise clear.
As a result, attacks similar to those for \mytt{LIKE} may be 
executed.

For instance, the condition
\mytt{WHERE right(name,5) = 'Murry'},
which strips all characters but the last five,
matches both Murry and McMurry. As another example, if a first
name column contained multiple Paul's but only one Paula, then both
\mytt{substring(name FROM 0 FOR 4)='paul'} and \mytt{WHERE left(name,4)='Paul'}.
can be used to match both names.

For conditions containing these functions,
an extra noise layer is created for the
function. For example, given the condition
\mytt{WHERE right(name,5)<>'Murry'}, the noise layers that would
normally be created from the condition
\mytt{WHERE name$<>$'Murry'} would be created, and in addition
a dynamic noise layer specifically associated with the \mytt{right} function
would be created.

This extra dynamic noise layer has the following seed components:

\begin{itemize}
\item The function name
\item The constant (or constants, in the case of substring) associated with the function (i.e. the number 5 in the case of \mytt{right(name,5)})
\item The symbol \mytt{:$<>$} if a negative condition
\end{itemize}

In the case of \mytt{trim}, we use \mytt{ltrim}, \mytt{rtrim}, or \mytt{btrim} as the function name depending on whether the \mytt{trim} is \mytt{LEADING}, \mytt{TRAILING}, or \mytt{BOTH}.

In the case of \mytt{substring(col FOR X)} or \mytt{substring(col FROM 0 FOR X)}, then we treat it as the equivalent \mytt{left(col, X)} with respect to seeding.

The constant for the \mytt{trim}'s needs to be normalized so that the analyst can't add chaff and average out the noise by for instance scrambling the characters or adding extra redundant characters. To do this, we remove redundant
characters and put the characters in alpha-numeric order.

\subsection{String functions lower and upper}
\label{sec:upperLower}

The current implementation also supports the string functions
\mytt{lower} and \mytt{upper}. While these to our knowledge cannot be used
in a difference attack, without additional noise layers they could
in many cases be used in a limited chaff attack to obtain three 
static noise samples
for the same query. For instance, suppose that the first letter
of last names is capitalized. The analyst could obtain three noise
samples with the following three conditions:

\begin{footnotesize}
\begin{verbatim}
WHERE name <> 'Smith'
WHERE upper(name) <> 'SMITH'
WHERE lower(name) <> 'smith'
\end{verbatim}
\end{footnotesize}

When \mytt{upper} or \mytt{lower} appears in a positive equality or
with the \mytt{IN} comparator, then nothing
special is required to seed the associated noise layers. The
column is floated (prior to the case change) and handled as usual.

When they appear in a condition that does not float (negative
equality or \mytt{NOT IN}) then an
extra static noise layer is generated. This noise layer is
the same regardless of whether the analyst executes \mytt{upper}
or \mytt{lower}: diffix-birch forces the string to lower case for the
purpose of seeding the noise layer. As such,  seed components include:

\begin{footnotesize}
\begin{verbatim}
[lower(value), ':<>', ':lower']
\end{verbatim}
\end{footnotesize}

where \mytt{lower(value)} is the string in lower case, and \mytt{:lower} is
an extra symbol.

\subsection{Functions that produce ranges}

There are a number of functions that, when combined with a
condition, effectively produce ranges.
For instance, Figure~\ref{fig:rangeFuncs} show how \mytt{trunc} and
\mytt{year} mimic the corresponding ranges specified with \mytt{BETWEEN}.
These include all of the datetime functions in
Appendix~\ref{app:datetime}, and the math functions
\mytt{trunc} and \mytt{round}.

\begin{figure}
\center
\begin{footnotesize}
  \begin{tabular}{l | l}
\hline

SELECT count(*) &
SELECT count(*) \\
FROM table &
FROM ( \\
WHERE age &
\hspace{0.3cm} SELECT uid,  \\
\hspace{0.3cm} BETWEEN 10 AND 20 &
\hspace{0.6cm} trunc(age, -1) AS tr\_age \\
 & \hspace{0.3cm} FROM table) t \\
 & WHERE tr\_age = 10 \\
\hline
\hline
SELECT count(*) &
SELECT count(*) \\
FROM table &
FROM ( \\
WHERE date  &
\hspace{0.3cm} SELECT uid,  \\
\hspace{0.3cm} BETWEEN '2016-01-01' &
\hspace{0.6cm} year(date) AS yr \\
\hspace{0.3cm} AND '2016-12-31' &
\hspace{0.3cm} FROM table) t \\
 & WHERE yr = 2016 \\

\hline
  \end{tabular}
\end{footnotesize}
\caption{A variety of math and datetime functions can group rows
as ranges. Here the query on the left and right of each pair
produce the same results (using \mytt{trunc} and \mytt{year}
respectively).
}
\label{fig:rangeFuncs}
\end{figure}

Without getting into details, our implementation recognizes 
these functions and associated conditions as requiring the
same noise layers as their explicit counterparts. Note that
all of the functions listed above are "naturally" snapped.
\mytt{trunc} and \mytt{round} operate on powers of 10 (0.01, 0.1, 1, 10, ...),
and with the exception of \mytt{weekday} and \mytt{quarter},
the datetime functions operate on the same natural
datetime boundaries as diffix-birch's snapped alignment.
Note that strictly speaking there are other math
functions like \mytt{floor} and \mytt{ceil}, and even casting a \mytt{real}
to an \mytt{int}, that also can generate ranges. These, however,
only operate at the level of single integers, and so we feel
provide inadequate expressiveness from which to effectively
be used for attacks.

\subsection{Function concat}

Diffix-birch allows the function \mytt{concat}.
\mytt{concat} can be used to mimic \mytt{AND}'d positive conditions.
For instance, the two queries in Figures~\ref{fig:concat} produce
the same buckets.

\begin{figure}
\center
\begin{footnotesize}
  \begin{tabular}{l | l}
\hline

SELECT bday,age, &
SELECT concat(c\_bday, '-', c\_age), \\
\hspace{0.6cm} count(*) &
\hspace{0.6cm} count(*) \\
FROM table &
FROM (SELECT uid, \\
GROUP BY bday,age &
\hspace{0.3cm} cast(bday, text) AS c\_bday, \\
 & \hspace{0.3cm} cast(age, text) AS c\_age \\
 & \hspace{0.3cm} FROM table) t \\
 & GROUP BY concat(c\_bday, '-', c\_age) \\

\hline
  \end{tabular}
\end{footnotesize}
\caption{The \mytt{concat} function can be used to mimic positive \mytt{AND}'d
conditions. The above queries produce identical buckets. 
}
\label{fig:concat}
\end{figure}

Diffix-birch floats the
columns used by \mytt{concat} and generates noise layers identical to
those produced by positive \mytt{AND}'d conditions.

\subsection{Noise reporting functions}

In order to help the analyst gauge the accuracy of query results,
diffix-birch provides a number of functions that report the amount
of the added noise (the standard deviation).
Especially for functions that report
sums or the count of rows, the amount of noise may vary
substantially since it is proportional to the contribution of
some of the largest users. These functions are \mytt{count\_noise},
\mytt{sum\_noise}, \mytt{avg\_noise}, and \mytt{stddev\_noise}.

Since the noise amount is itself based on values from multiple
distinct users, we do not expect it to be attackable.
Never-the-less, to stay on the safe side,
the reported noise is a rounded value. 
The unit of rounding increases with the size of the standard
deviation such that the rounded value is within roughly
5\% of the true value.

\bibliographystyle{acm}
\bibliography{ref}

\clearpage

\appendix

\section{Complete Diffix-birch SQL Syntax}
\label{sec:syntax}

\begin{footnotesize}
\begin{verbatim}
SELECT [DISTINCT]
  field_expression [, ...]
  FROM from_expression [, ...]
  [ WHERE where_expression [AND ...] ]
  [ GROUP BY column_expression | position [, ...] ]
  [ HAVING having_expression [AND ...] ]
  [ ORDER BY column_name [ASC | DESC] | position [, ...] [ LIMIT amount ] [ OFFSET amount ] ]

field_expression :=
  * | table_name.* | column_expression [AS alias]

column_expression :=
  [table_name.]column_name |
  aggregation_function([DISTINCT] column_name) |
  function(column_expression) |
  column_expression binary_operator column_expression |
  column_expression::data_type

binary_operator :=
  + | - | * | / | ^ | %

data_type :=
  integer | real | text | boolean | datetime | date | time

from_expression :=
  table | join

table :=
  table_name [[AS] alias] | (select_expression) [AS] alias

join :=
  table CROSS JOIN table |
  table { [INNER] | { LEFT | RIGHT } [OUTER] } JOIN table ON where_expression

aggregation_function :=
  COUNT | SUM | AVG | MIN | MAX | STDDEV | MEDIAN

where_expression :=
  column_expression equality_operator (value | column_expression) |
  column_expression inequality_operator (numerical_value | datetime_value) |
  column_expression BETWEEN value AND value |
  column_expression IS [NOT] NULL |
  column_expression [NOT] IN (constant [, ...])
  column_expression [NOT] LIKE | ILIKE string_pattern [ESCAPE escape_string]

having_expression :=
    column_expression comparison_operator (value | column_expression)

comparison_operator :=
    equality_operator | inequality_operator

equality_operator :=
    = | <>

inequality_operator :=
    > | >= | < | <=
\end{verbatim}
\end{footnotesize}

\clearpage

\section{Diffix-birch Supported Functions}

\subsection{String Functions}

\begin{itemize}
\item BTRIM()
\item CONCAT()
\item HEX()
\item LEFT()
\item LENGTH()
\item LOWER()
\item LCASE()
\item LTRIM()
\item RIGHT()
\item RTRIM()
\item SUBSTRING()
\item TRIM()
\item UPPER()
\item UCASE()
\end{itemize}

\subsection{Datetime Functions}
\label{app:datetime}

\begin{itemize}
\item year()
\item quarter()
\item month()
\item day()
\item hour()
\item minute()
\item second()
\item weekday()

\item EXTRACT()
  \begin{itemize}
    \item year
    \item quarter
    \item month
    \item day
    \item hour
    \item minute
    \item second
  \end{itemize}

\item DATE\_TRUNC()
  \begin{itemize}
    \item year
    \item quarter
    \item month
    \item day
    \item hour
    \item minute
    \item second
  \end{itemize}
\end{itemize}

\subsection{Math Functions}
\label{app:math}

Operators +, -, *, /, \^, and \%.

\begin{itemize}
\item ABS()
\item BUCKET()
\item CEIL()
\item DIV()
\item FLOOR()
\item MOD()
\item POW()
\item ROUND()
\item SQRT()
\item TRUNC()
\end{itemize}

\subsection{Noise Indicators}

\begin{itemize}
\item count\_noise()
\item sum\_noise()
\item avg\_noise()
\item stddev\_noise()
\end{itemize}

\subsection{Casting}

The following tables indicate what casting combinations are allowed.

\begin{center}
\begin{tabular}{| l | c | c | c | c |}
\hline
\multicolumn{5}{| c |}{Casting} \\
\hline
\hline
from/to & text & integer & real & boolean \\
\hline
text & X & X & X & X \\
integer & X & X & X & X \\
real & X & X & X & X \\
boolean & X & X & X & X \\
date & X &   &   &   \\
time & X &   &   &   \\
datetime & X &   &   &   \\
interval & X &   &   &   \\

\hline
\end{tabular}
\end{center}

\begin{center}
\begin{tabular}{| l | c | c | c | c |}
\hline
\multicolumn{5}{| c |}{Casting} \\
\hline
\hline
from/to & date & time & datetime & interval \\
\hline
text & X & X & X & X \\
integer &   &   &   &   \\
real &   &   &   &   \\
boolean &   &   &   &   \\
date & X &   &   &   \\
time &   & X &   &   \\
datetime & X & X & X &   \\
interval &   &   &   & X \\

\hline
\end{tabular}
\end{center}

\end{document}